\documentclass[aps,prb,twocolumn]{revtex4}
\usepackage{bm,color,amsmath,amssymb,mathrsfs,latexsym,graphicx,psfrag}

\newcommand{\bk}{{\mathbf k}}

\newcommand{\be}{\begin{equation}}
\newcommand{\ee}{\end{equation}}

\def\be{\begin{equation}}
\def\ee{\end{equation}}
\def\bea{\begin{eqnarray}}
\def\eea{\end{eqnarray}}

\begin{document}
\title{Topological nodal line semimetals}
\author{Chen Fang}%
\affiliation{Beijing National Laboratory for Condensed Matter Physics and Institute of Physics, Chinese Academy of Sciences, Beijing 100190, China}%

\author{Hongming Weng}%
\affiliation{Beijing National Laboratory for Condensed Matter Physics and Institute of Physics, Chinese Academy of Sciences, Beijing 100190, China}%
\affiliation{Collaborative Innovation Center of Quantum Matter, Beijing, China}

\author{Xi Dai}%
\affiliation{Beijing National Laboratory for Condensed Matter Physics and Institute of Physics, Chinese Academy of Sciences, Beijing 100190, China}%
\affiliation{Collaborative Innovation Center of Quantum Matter, Beijing, China}

\author{Zhong Fang}%
\affiliation{Beijing National Laboratory for Condensed Matter Physics and Institute of Physics, Chinese Academy of Sciences, Beijing 100190, China}%
\affiliation{Collaborative Innovation Center of Quantum Matter, Beijing, China}

\date{\today}
\begin{abstract}
We review the recent, mainly theoretical, progress in the study of topological nodal line semimetals in three dimensions. In these semimetals, the conduction and the valence bands cross each other along a one-dimensional curve in the three-dimensional Brillouin zone, and any perturbation that preserves a certain symmetry group (generated by either spatial symmetries or time-reversal symmetry) cannot remove this crossing line and open a full direct gap between the two bands. The nodal line(s) is hence topologically protected by the symmetry group, and can be associated with a topological invariant. In this Review, (i) we enumerate the symmetry groups that may protect a topological nodal line; (ii) we write down the explicit form of the topological invariant for each of these symmetry groups in terms of the wave functions on the Fermi surface, establishing a topological classification; (iii) for certain classes, we review the proposals for the realization of these semimetals in real materials and (iv) we discuss different scenarios that when the protecting symmetry is broken, how a topological nodal line semimetal becomes Weyl semimetals, Dirac semimetals and other topological phases and (v) we discuss the possible physical effects accessible to experimental probes in these materials.
\end{abstract}
\maketitle

\section{Introduction}

Topological semimetals (TSM) are defined as systems where the conduction and the valence bands cross each other in the Brillouin zone (BZ), and the crossing is non-accidental, \emph{i. e.}, cannot be removed by perturbations on the Hamiltonian without breaking any of its symmetries. If there be no symmetry, two bands, when close in energy, will hybridize with each other and maintain a gap in-between, through a mechanism known as the band repulsion; however, when in the presence of certain symmetries (\emph{e. g.} crystalline symmetries and time-reversal symmetry), the two crossing bands may have different quantum numbers such that they cannot be hybridized. From this, we see that all band crossings, hence all topological semimetals, can only be protected by symmetries and hence belong to symmetry protected topological phases of matter.

In three dimensions (3D),  two bands can cross each other either at discrete points or along a closed curve. In the former case, there are Weyl semimetals\cite{Murakami2007,Wan2011} and Dirac semimetals\cite{Wang2012} that have been intensively studied in theory as well as in experiment. In the latter case, the curve where the bands cross is called a nodal line\cite{Burkov_Topological_nodal_semimetals_2011PRB}, which may either take the form of an extended line running across the BZ, whose ends meet the at the BZ boundary\cite{Chen2015}, or wind into a closed loop inside the BZ\cite{XuGang_HgCrSe_2011_PRL}, or even form a chain consisting of several connected loops (nodal chain)\cite{nodal_chain}. Topological semimetals with such line band crossings are called topological nodal line semimetals (TNLSM). In principle, one may have TNLSM in both quasi-2D\cite{Lin2016} and 3D systems\cite{Chiu2014,Yang2016}, and in this Review, we will focus on the latter.

In a topological gapped phase, such as Chern insulator, topological insulator or topological crystalline insulator, the nontrivial topology of the bands can be characterized by a topological invariant, a quantum number that depends on the Bloch wave functions of the valence bands as a whole. The form of the topological invariant is determined by only two factors: dimension and symmetry. Similarly for a topological semimetal, one can also associate with each band crossing (either a point node or a line node) a topological invariant, whose form depends on the symmetry group that protects the nodal structure. Given the dimension of the node and the symmetry group that protects it, one or several topological invariants can be found, and these quantum numbers provide a full topological classification of the nodes. Up to this day, the classification of TNLSM is yet to be complete. Theoretically, people have found: mirror reflection protected nodal lines with a $Z$-invariant\cite{XuGang_HgCrSe_2011_PRL,TlTaSe2,PbTaSe2,CaAgX,ZrTe}, nodal line protected by inversion, time-reversal and spin rotation symmetries with two $Z_2$-invariants\cite{allcarbon_nodeLine2014,Ca3P2,Ca3P2-2,Cu3NPd_Kane, Cu3NPd_Yu,LnX,Murakami_CaSrYb, ChenXQ_Be,CaTe,BaSn2,BP_Zhao2015,CaP3,carbon_fanzhang}, screw rotation protected double nodal lines (to be defined later) with a $Z$-invariant\cite{Carter2012,CFang_NLSM_PRB,BaMX3}, et cetera.

While a topological classification tells us how many topologically different types of nodes exist in systems with a given symmetry group, only numerical calculation, mainly first principles calculations that compute the band structure and Bloch wave functions, can compute this invariant in a given compound and determine if it is a TSM or not. The calculation has proved more challenging than that of the topological invariant in a gapped system, because (i) the slow convergence in a gapless system and (ii) the band crossing point is not always at a high-symmetry point, along a high-symmetry line or even on a high-symmetry plane. We will review numerical efforts that have resulted in the proposals of various materials systems as TNLSM protected by different symmetry groups.

When the protecting symmetry is broken in a TNLSM, the nodal line is either fully gapped or gapped into several nodal points. For example, without spin-orbital coupling (SOC), TaAs was\cite{TaAs_Weng,HuangSM_Weyl}, in first principles calculation, shown to be a TNLSM protected by mirror reflection and spin-rotation symmetries with two nodal lines, and when SOC was turned on, each nodal line is gapped into three pairs of Weyl nodes. Another example is the double nodal line in SrIrO$_3$, which is gapped into a pair of non-symmorphic Dirac nodes when a certain mirror reflection symmetry is broken\cite{Carter2012,Fang2016}. Therefore, understanding how a nodal line evolves upon symmetry breaking can help predict new topological materials.

Unlike most topological phases, TNLSMs in general do \emph{not} have \emph{protected} boundary modes\cite{CFang_NLSM_PRB}. Therefore, identifying them in experiments poses a challenge to the experimentalists. Angle resolved photoemission spectroscopy (ARPES) has been used to resolve the nodal band structure in the bulk\cite{PbTaSe2,TlTaSe2}, but the limited momentum resolution in the perpendicular direction prevents these efforts from being deterministic. Quantum oscillation measurements\cite{Hu2016} were done to map the Fermi surface of ZrSiSe and ZrSiTe as well as the Berry phase along a closed loop on the Fermi surface, partially supporting the proposal of TNLSMs in these materials. Another unanswered question is the fate of these materials in the presence of electron correlation. It has been proposed that the screening effect is qualitatively different from that in normal metals\cite{Huh2016}, and that in the presence of strong electron interaction, the quantum phase transition from a TNLSM to a nodal point semimetal or a gapped system belongs to a new universality class\cite{Han2016}.

Below is the outline of the Review. In Section II, we briefly go through the topological classification of TNLSM protected by several symmetry groups by writing down the expression of the topological invariants in terms of the Bloch wave functions. In Section III, we review the several materials proposed by first principle calculations to be TNLSMs. In Section IV, we discuss the various scenarios of how the nodal line evolves when the protecting symmetry is broken and in Section V we discuss the experimental consequences and the many-body effects in TNLSMs so far proposed in literature.

\section{Topological classification}

Topological invariant gives a quantitative description of the topology of a system. Let us first review its definition in a `gapped' band structure, where `gapped' means that at each momentum $\bk$ in BZ, there is a finite direct gap between the conduction and the valence bands, while the indirect gap is allowed to close. Given two Hamiltonians $H_1$ and $H_2$, if $H_1$ can be tuned to $H_2$ without (i) closing the gap or (ii) breaking a given symmetry group, then $H_1$ and $H_2$ are said to be \emph{topologically equivalent} under this symmetry group. By this equivalence, one can divide the Hamiltonians into different equivalent classes, and there a one-to-one isomorphism mapping each class to a set of integers: the explicit form of this isomorphism is the topological invariant(s) protected by the symmetry group. Well known examples of topological invariants include: Chern numbers in 2D Chern insulators (quantum anomalous Hall states)\cite{Haldane1988}, $Z_2$ indices in 2D and 3D insulators protected by time-reversal and charge conservation\cite{Kane2005,Fu2007a,Moore2007}, the spin Chern numbers in 2D quantum spin Hall states protected by spin rotation about the $z$-axis\cite{Kane2005a,Bernevig2006}, et cetera.

For TSM, the definition of topological invariant needs modification as we cannot have a well-defined conduction or valence bands throughout the BZ: at some point the two bands cross each other. Given a nodal structure, say a point node, in BZ, we first use an imaginary manifold to enclose without touching the node. On the enclosing manifold, the conduction and the valence bands do not touch each other, having a full direct gap. Therefore, topological invariants can be defined on the enclosing manifold and we identify this invariant as the topological invariant of the node inside. We use the example of Weyl point to illustrate this process\cite{Murakami2007}, which is the same one used in numerics to calculate the monopole charge of a Weyl point\cite{Wan2011,TaAs_Weng}. Given a Weyl point, we consider a sphere in k-space to enclose this point. (Here one needs to make sure that only one band crossing point is inside.) Then since the bands are `gapped' on the surface of the sphere, we can calculate its Chern number. When the Chern number is $\pm1$, we know that the Weyl point has monopole charge of $\pm1$; if the Chern number is $\pm2$, we know that the nodal point is actually a double-Weyl point\cite{Fang2012} with monopole charge of $\pm2$. Another example we use is the Dirac point protected by $C_{4v}$  (existing in \emph{e. g.} Cd$_3$As$_2$\cite{Wang2013}). Here since the Dirac point is only allowed to move along the $k_z$-axis in the BZ, we consider two points $p_{1,2}$ along the $k_z$-axis above and below the Dirac point, to enclose the Dirac point. At $p_1$ and $p_2$, the conduction and the valence bands are separated, and each band is double degenerate with $C_4$ eigenvalues $e^{\pm{i}\pi/4}$ or $-e^{\pm{i}\pi/4}$. At $p_{1,2}$ we count the number of valence bands that have $C_4$-eigenvalue $e^{\pm{i}\pi/4}$, denoted by $N_{1,2}$ respectively. The topological invariant of the Dirac point is given by $z=N_1-N_2$.

For TNLSMs, we have three types of enclosing manifolds, being zero-dimensional, one-dimensional and two-dimensional, respectively. If the nodal line is fixed by symmetry (usually a mirror reflection symmetry) to some high-symmetry plane, we choose (a) two points on the same plane, on different sides of the nodal line, respectively [see Fig.\ref{fig:2}(a)]. If the nodal line is not fixed to any high-symmetry plane, then we consider the two following enclosing manifolds: (b) a loop that links with the nodal line [see Fig.\ref{fig:2}(b)] and (c) a sphere or torus that encloses the nodal line [see Fig.\ref{fig:2}(c)].

\subsection{Nodal lines protected by mirror reflection symmetries}

In real space, a mirror reflection symmetry can be defined as
\bea\label{eq:0}
M: (x,y,z)\rightarrow(x,y,-z),
\eea
whereas in momentum space
\bea
M: (k_x,k_y,k_z)\rightarrow(k_x,k_y,-k_z).
\eea
The symmetry can be represented by a unitary operator acting on one-electron wave functions, satisfying
\bea\label{eq:1}
M^2=(-1)^{2S},
\eea
where $S$ is the spin of a single particle. From Eq.(\ref{eq:1}) we see an important distinction between spinless ($S=0$) and spinful ($S=1/2$) particles. For the former, $M$ has eigenvalues $\pm1$ and for the latter, $\pm{i}$. Physically, this factor of $i$ is due to the fact that in a spin-orbit coupled system, the reflection also acts on the spin degrees of freedom. Mark that spinful systems can be viewed as spinless systems when the full SU(2) spin-rotation symmetry is preserved, because the spatial and the spin degrees of freedom are decoupled. When a non-interacting system described by Hamiltonian $H(k_x,k_y,k_z)$ has mirror reflection symmetry $M$, we have
\bea\label{eq:2}
MH(k_x,k_y,-k_z)M^{-1}=H(k_x,k_y,k_z).
\eea
Note that $k_z\rightarrow-k_z$ on the left is due to that the mirror operation also flips the $z$-component of the momentum. By Eq.(\ref{eq:2}), we see that at the two high-symmetry planes in the BZ, namely $k_z=0$ and $k_z=\pi$, the mirror operator and the Hamiltonian have the same eigenstates, such that we can use the eigenvalues of $M$ to label the bands on these two planes. Without other symmetries present, all bands are generically non-degenerate, and suppose there are two bands labeled by $+1$ and $-1$ respectively (assuming $S=0$), then these two bands are disallowed by mirror symmetry to hybridize with each other. Therefore, two bands with opposite mirror eigenvalues can cross each other at $\bk$-points satisfying
\bea\label{eq:3}
E_+(\bk)=E_-(\bk).
\eea
Since $\bk$ has two free components ($k_z$ being fixed to $0$ or $\pi$), in Eq.(\ref{eq:3}) we have two variables to satisfy one equation, meaning that the solution space is generically one-dimensional, \emph{i. e.}, a nodal line. The same discussion follows when $S=1/2$. We emphasize that the two bands can cross only at $k_z=0$ and $k_z=\pi$ planes: away from them, there is no quantum number to distinguish the bands or prevent hybridization.
\begin{figure}
\includegraphics[width=8cm]{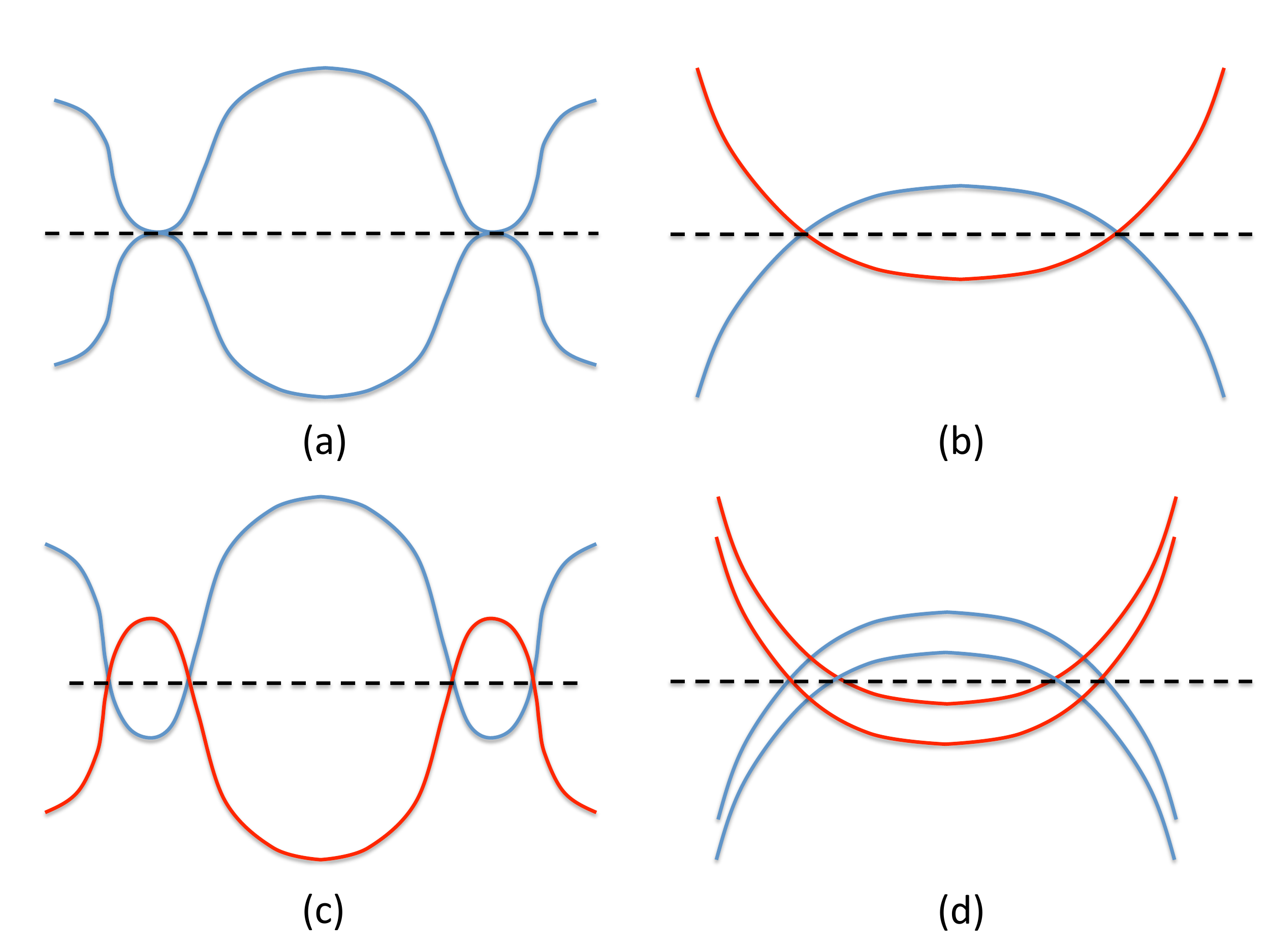}
\caption{\label{fig:1}Dispersion along a cut across the nodal lines protected by mirror reflection symmetries carrying different topological invariants. Different colors represent opposite mirror eigenvalues. (a) An accidental nodal line which has $\zeta_0=0$. (b) A protected nodal line carrying $\zeta_0=1$. (c) Two nodal lines (one inner and one outer) carrying opposite invariants. (d) Two nodal lines carrying the same invariant.}
\end{figure}
Since the nodal line is fixed to the high-symmetry planes by mirror symmetry, we use the zero-dimensional enclosing manifold. On the two sides of the nodal line we pick two points $p_1$ and $p_2$. At $p_{1,2}$, the conduction and the valence bands are separated in energy, and one can count the number of bands below the Fermi energy that have mirror eigenvalue of $+1$, denoted by $N_{1,2}$. The topological invariant is then given by
\bea
\zeta_0=N_1-N_2.
\eea
$\zeta_0=0$ corresponds to the case shown in Fig.\ref{fig:1}(a), and it means that the crossing is only accidental and can be removed without breaking the mirror symmetry; $\zeta=1$ corresponds to the case shown in Fig.\ref{fig:1}(b), and it means that the crossing is between two bands that have opposite mirror eigenvalues. $\zeta=2$ corresponds to the case shown in Fig.\ref{fig:1}(c), and it means that there are two nodal lines resulting from two pairs of bands with opposite eigenvalues. Here we remark that the 0D invariant $\zeta_0$ is a $Z$-invariant, not $Z_2$-invariant. Suppose by tuning the parameters we can put two nodal lines at the same $\bk$-point in BZ, then if the invariant is $Z_2$, the two nodal lines will `cancel' each other and create a full gap, but if the invariant is $Z$, it depends on whether the two lines have the same or opposite invariants. If the invariants are the same, then the two lines will not gap each other (as in Fig.\ref{fig:1}(d)), and if the invariants are opposite, putting these nodal lines together will create a full gap.

\subsection{Nodal lines protected by inversion, time-reversal and SU(2) spin-rotation symmetries}

Here we first assume that all the three symmetries are present in our system. Since SU(2) is a symmetry, we can redefine time-reversal operator, combining it with a $\pi$ spin rotation about the $y$-axis
\bea
T\rightarrow{T}e^{is_y\pi},
\eea
after which we have $T^2=+1$ instead of $-1$ for fermions. Since both inversion, $P$, and $T$ reverse the momentum $\bk\rightarrow-\bk$, $P*T$ is an anti-unitary symmetry that preserves the momentum. Since $[P,T]=0$, we have
\bea\label{eq:4}
(P*T)^2=P^2T^2=1.
\eea
Eq.(\ref{eq:4}) dictates that it can be represented as
\bea
P*T=K
\eea
where $K$ is complex conjugation, in a proper orbital basis. In this basis, $P*T$-symmetry ensures that
\bea
H(\bk)=H^\ast(\bk),
\eea
or that $H(\bk)$ is real at each $\bk$.

Away from the crossing lines, the Hamiltonian can be `flattened' as
\bea\label{eq:5}
Q(\bk)=I-\sum_{n\in{occ.}}|u_{n}(\bk)\rangle\langle{u}_n(\bk)|.
\eea
From Eq.(\ref{eq:5}), we see that the eigenfunctions of $Q(\bk)$ are the same as those of $H(\bk)$, but eigenvalues of $Q(\bk)$ are either zero or one, depending on whether $n$ is an occupied or unoccupied band. Then, we note that on any compact manifold that does not contain any crossing point, the Hamiltonian $H(\bk)$ can always be deformed into $Q(\bk)$ without breaking any symmetry or closing the gap, \emph{i. e.}, $H(\bk)$ and $Q(\bk)$ have the same topology.

Before going further, a brief review the concept of homotopy groups in algebraic topology is due. Consider continuous mappings from an $n$-sphere $S^n$ to a space $M$ (in this case the space of all occupied bands). If two such mappings, $\phi_1$ and $\phi_2$, can be continuously deformed into each other, then we say that $\phi_1$ and $\phi_2$ are equivalent to each other. Mappings that are equivalent to each other form an equivalent class, and each class correspond to an element in the homotopy group, denoted by $\pi_n(M)$. For example, if the homotopy group has only one element, $\pi_n(M)=\{e\}$, then we know that all mappings from $S^n$ to $M$ are equivalent to each other, and therefore equivalent to a trivial mapping where all points in $S^n$ map to the same point in $M$. If the homotopy group is $Z_2$, $\pi_n(M)=Z_2$, then we know that all mappings are either equivalent to a trivial mapping or to a nontrivial mapping. This mathematical definition can be paraphrased in our physical context: $S^{n\ge1}$ is our enclosing manifold; $M$ is the Hilbert space spanned by all occupied bands; the Hamiltonian $H(\bk)$ is the mapping; and the condition that two mappings can continuously deform into each other corresponds to the case where $H_1(\bk)$ can be adiabatically transformed to $H_2(\bk)$ without gap closing. Therefore, the homotopy group of $Q(\bk)$ exactly gives the topological classification of the nodal line enclosed by the $n$-sphere.

\begin{figure}
\includegraphics[width=8cm]{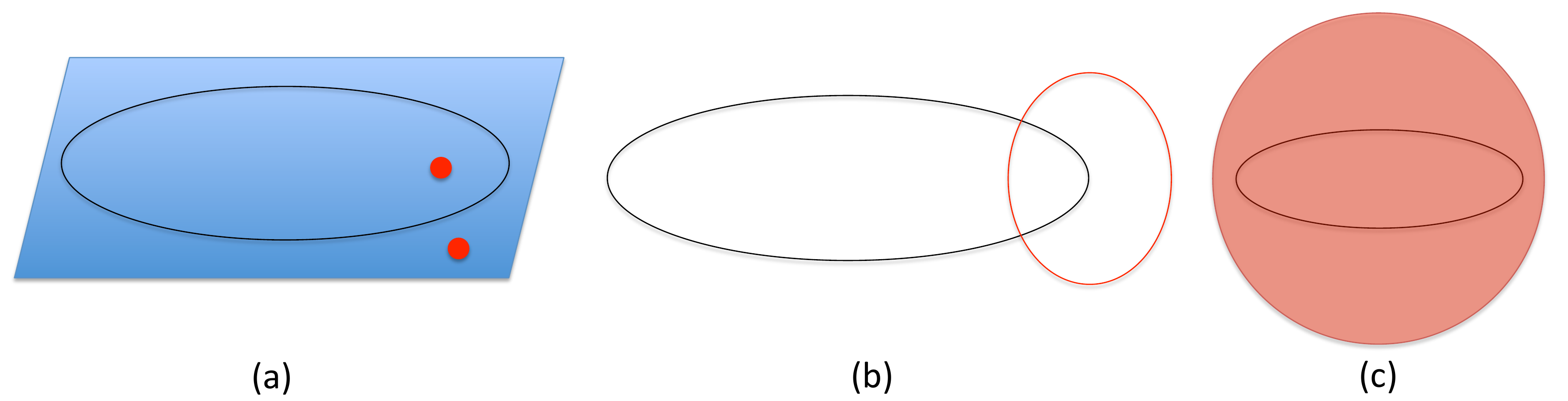}
\caption{\label{fig:2}Manifolds of different dimensions ($S^0$, $S^1$ and $S^2$) that enclose a nodal line: (a) Two points ($S^0$) inside and outside the nodal line pinned to (protected by) a mirror plane, (b) a loop ($S^1$) that threads the nodal line and (c) a sphere ($S^2$) surrounding the entire nodal line.}
\end{figure}

If $Q(\bk)$ is real, it is an element of the real Grassmanian manifold, or
\bea
Q(\bk)\in\frac{O(m+n)}{O(m)\oplus{O}(n)}.
\eea
It is known that the homotopy groups of this manifold are
\bea\label{eq:6}
\pi_1(\frac{O(m+n)}{O(m)\oplus{O}(n)})=\pi_2(\frac{O(m+n)}{O(m)\oplus{O}(n)})=Z_2
\eea
for $m,n>2$.
Eq.(\ref{eq:6}) means that if we enclose the nodal line with either a ring or a sphere, the topological classification of the wave functions on the ring/sphere is $Z_2$. This means that for a nodal ring, we have two \emph{independent} $Z_2$-indices (denoted by $\zeta_1$ and $\zeta_2$, defined on a ring that links with the line (Fig.\ref{fig:2}(b)) and on a sphere that encloses the whole line (Fig.\ref{fig:2}(c)). 
If $\zeta_1$ is zero, we infer that the line crossing is purely accidental and can be removed by an arbitrarily small perturbation that preserves all symmetries. Therefore, all topological nodal rings protected by this symmetry group must all have $\zeta_1=1$. The explicit expression of $\zeta_1$ is simply the Berry phase on the ring that links with the nodal line:
\bea
(-1)^{\zeta_1}=\oint{dk}\mathbf{A}(\bk)\cdot{d}\bk,
\eea
where
\bea
\mathbf{A}(\bk)=-i\sum_{n\in{occ.}}\langle{u_n}(\bk)|\partial_\bk|u_n(\bk)\rangle
\eea
is the Abelian Berry connection. It can be proved that when $H(\bk)$ is real, the Berry phase associated with any loop must be quantized to either $0$ or $\pi$. If it is zero, then we can smoothly shrink this loop to a single point; but if it is $\pi$, the loop cannot shrink to a point, as an infinitesimal loop necessarily has zero Berry phase. In the latter case, there must be a point inside the loop where the Berry phase cannot be defined, that is, where the conduction and the valence bands cross.

\begin{figure}
\includegraphics[width=8cm]{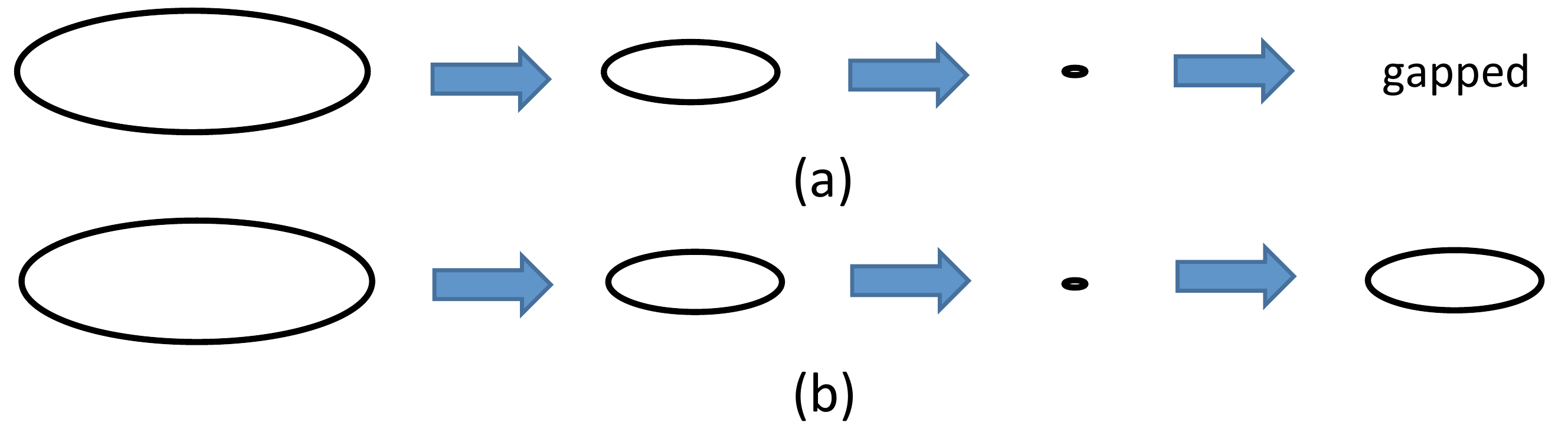}
\caption{\label{fig:3}(a) By tuning $m$ from positive to negative in Eq.(\ref{eq:H2}), the nodal line with $\zeta_2=0$ is fully gapped and (b) by tuning $m$ from positive to negative in Eq.(\ref{eq:model2}), the nodal line with $\zeta_2=1$ first shrinks to a point but grows into a line again.}
\end{figure}

We have understood that $\zeta_1=0$ and $\zeta_1=1$ means that the nodal ring is accidental and protected, respectively. What is the physical meaning of the second index? When the second index $\zeta_2$ is zero, then although the nodal line is stable agains perturbations, the nodal line may still shrink to a point and be gapped by a continuous tuning of the Hamiltonian. This point can be illustrated by the following example. Consider an effective Hamiltonian near $\Gamma$
\bea\label{eq:H2}
H(\bk)=(m-k^2)\sigma_z+k_z\sigma_x,
\eea
where $P=\sigma_z$ and $T=K\sigma_z$ and of course $P*T=K$ as promised. If $m>0$, the two bands cross each other on the $k_z=0$-plane, making a nodal circle of radius $r=\sqrt{m}$. It is obvious that as $m$ decreases, the nodal circle shrinks, and at $m=0$, it shrinks into a single point at $\bk=0$, and when we further decrease $m$ to $m<0$, the nodal circle vanishes [see Fig.\ref{fig:3}(a)]. This is a typical example when the second index of a nodal ring is zero.
When $\zeta_2=1$, it means that on the surface of the sphere that surrounds the nodal line, the periodic part of the Bloch wave functions cannot be adiabatically tuned to the same function and also means that one cannot shrink the sphere to a single point without meeting a singularity in the process. In this case, the nodal line cannot shrink to a point and be gapped out, which can be proved by contradiction: suppose the nodal line can vanish by tuning the Hamiltonian, one can first gap out the nodal line by shrinking it, such that there is no singularity inside the sphere, then one can shrink the sphere to a single point, contradicting the assumption that $\zeta_2=1$. We use the following example to illustrate a nodal line with $\zeta_2=1$.
\bea\label{eq:model2}
H(\bk)=k_xs_x+k_y\tau_ys_y+k_zs_z+m\tau_xs_x,
\eea
where $\tau_i$ and $s_i$ are Pauli matrices acting on two isospin degrees of freedom. The spectrum is given by
\bea
E(\bk)=\pm\sqrt{k_z^2+(\sqrt{k^2_x+k^2_y}\pm{m})^2}.
\eea
The band crossing can be found by solving $E(\bk)=0$, yielding $k_z=0$ and $\sqrt{k_x^2+k_y^2}=|m|$, i.e., a nodal line on the $xy$-plane of radius $\sqrt{|m|}$. As $m$ changes from positive to negative, the radius decreases and shrinks to zero at $m=0$ but increases again when $m$ becomes negative [see Fig.\ref{fig:3}(b)]. Therefore, this nodal line cannot shrink to a point and then be gapped out by tuning the parameters.

One can draw an analogy between the $\zeta_2$ of the nodal line and the monopole charge of a Weyl point, as both are defined on a sphere that encloses the nodal structure, with a key difference that here the monopole charge is a $Z_2$-charge in the former and $Z$-charge in the latter. This difference implies that when two nodal rings with $\zeta_2=1$ meet each other, they will necessarily annihilate each other, while two Weyl points with $C=1$ will not annihilate each other, but form a double Weyl point. The Nelson-Nanomiya theorem also applies to this $Z_2$-charge, which states that in a lattice model, the nodal lines with $\zeta_2=1$ must appear in pairs. This is another distinction between nodal lines with $\zeta_2=0$ and $\zeta_2=1$: while the ring can be annihilated or created in singles in the former case, in the latter case they can only be annihilated or created in pairs.

The explicit form of $\zeta_2$, in terms of the wave functions, is not very concise, and readers are encouraged to read the Appendices of Ref.[\onlinecite{CFang_NLSM_PRB}] for more information. Here we give an intuitive understanding of this invariant. Since the Hamiltonian $H(\bk)$ is real, we can find for each $\bk$-point a set of \emph{real} eigenfunctions of $H(\bk)$, denoted by $|u_n(\bk)\rangle\in{Real}$. Then, a natural question is whether there is a choice of $|u_n(\bk)\rangle$ that are smooth on the entire sphere that encloses the nodal line? If $\zeta_2=1$, it means that this `smooth, real gauge' does not exist.

Before closing this subsection, a few technical comments are due. (i) The first invariant $\zeta_1$ was first identified as a $Z$-invariant in literature\cite{Burkov_Topological_nodal_semimetals_2011PRB}, because the authors considered a two band model, and $\pi_1(O(2)/O(1)\oplus{O}(1))$ is indeed $Z$. Physically, it means that if there are only two bands, two nodal lines of the same charge meeting each other will not gap out, but once more bands are introduced, they can meet with the two crossing bands and gap out the nodal line. (ii) Another special case is $\pi_2(O(4)/O(2)\oplus{O}(2))=Z$, that is, when there are in total four bands and two conduction (valence) bands, the second index becomes a $Z$-index. This means that a nodal line may carry either positive or negative charges, and that two nodal lines with the same monopole charge will not annihilate each other, but when other bands are introduced into the model, the two nodal lines will be gapped when they meet each other.

\subsection{Double-nodal lines protected by twofold screw rotation, inversion and time-reversal}

Double-nodal lines appear when both the conduction and the valence bands are doubly degenerate and they cross each other along a nodal line, where the bands are fourfold degenerate. From its definition, we see that we need a symmetry such that the all bands are doubly degenerate, and then we need another symmetry to protect the band crossing. 

Even for spin-orbit coupled systems, $P*T$ makes sure that all bands are doubly-degenerate. In the presence of SOC, we cannot redefine $T$ such that $T^2=+1$, and therefore, we have
\bea
(P*T)^2=-1.
\eea
Since $P*T$ preserves the momentum and is an anti-unitary operator, we can prove that all bands are doubly degenerate. To be specific, for any Bloch state $|\psi_n(\bk)\rangle$, $P*T|\psi_n(\bk)\rangle$ must be an eigenstate at $\bk$ that is orthogonal to $|\psi_n(\bk)\rangle$.

Intuitively, it is understood that it is harder for two degenerate bands to cross each other than for non-degenerate bands. Consider an effective theory near the band crossing, then one needs at least a four-band model for the former and a two-band model for the latter. A four-band model is expanded in fifteen Dirac matrices (the identity matrix having nothing to do with band crossings) while a two-band model three Pauli-matrices. Then a double-line nodal requires the coefficients of fifteen Dirac matrices to vanish and a single nodal line only requires the coefficients of three Pauli matrices to vanish. Corroborating with this intuition is the fact that mirror reflection symmetry $M: (x,y,z)\rightarrow(x,y,-z)$ is insufficient to protect the crossing. 

For spinful systems, Eq.(\ref{eq:1}) gives that the mirror eigenvalues are either $+i$ or $-i$. Since $P*T$ commutes with $M$ and $P*T$ is anti-unitary, $|\psi_n(\bk)\rangle$ and $P*T|\psi_n(\bk)\rangle$ have opposite mirror eigenvalues, \emph{i. e.}, the two degenerate bands have opposite mirror eigenvalues. Therefore, when two degenerate bands, say band one and band two cross, the $+i$-subband in band one will repel the $+i$-subband in band two and similar repulsion exists between the $-i$-subbands in the two bands, creating a full gap as a result.

From this discussion, we may conjecture that in order for the two bands to have a protected crossing, we need the two subbands in a degenerate band to have the $\emph{same}$ quantum number of some symmetry. So far, one type of these symmetries have been found\cite{CFang_NLSM_PRB}: the twofold screw rotation, which acts in real space as
\bea\label{eq:8}
R: (x,y,z)\rightarrow(-x,-y,z-c/2)
\eea
where $c$ is the lattice constant in the $z$-direction. In the presence of inversion, the twofold screw rotation is equivalent to the following mirror plane $M'=P*R$
\bea
M': (x,y,z)\rightarrow(x,y,c/2-z).
\eea
It is easy to see that the only distinction between $M'$ and $M$ defined in Eq.(\ref{eq:0}) is that the mirror plane of $M'$ is located at $z=c/4$, while for $M$ it is at $z=0$. This offset of mirror plane leads to the following commutation relation between $P$ and $M'$
\bea
P*M'=T_{00\bar1}M'*P,
\eea
where $T_{00\bar1}$ is the translation along the $-z$-direction by one unit cell. At the BZ boundary $k_z=\pi$, $T_{00\bar1}=\exp(ik_zc)=-1$. Therefore, at $k_z=\pi$-plane, we have
\bea
\{P*T,M'\}=0.
\eea
Suppose on this plane, one subband of the conduction (or valence) bands have $M'=+i$, then the $M'$ eigenvalue of the other subband is found to be the same:
\bea
M'(P*T|+i\rangle)=-P*T(M'|+i\rangle)=-P*T(+i|+i\rangle)=+i(P*T|+i\rangle),
\eea
where we have used the fact that $P*T$ anti-commute with both $M'$ and $+i$. The same steps show that if one subband has $M'=-i$, the other subband also has $M'=-i$. When the degenerate conduction and the valence bands have opposite $M'$-eigenvalues on the $k_z=\pi$ plane, on this plane, the effective four-band Hamiltonian reads
\bea
H= \left(\begin{matrix} 
      \epsilon_v(k_x,k_y,\pi) & 0 & 0 & 0 \\
      0 & \epsilon_v(k_x,k_y,\pi) & 0 & 0 \\
      0 & 0 & \epsilon_c(k_x,k_y,\pi) & 0\\
      0 & 0 & 0 & \epsilon_c(k_x,k_y,\pi)
   \end{matrix}\right).
\eea
To understand why it takes such a simple form, we first notice that $M'$ symmetry requires the off-diagonal block to vanish, and $P*T=K(is_y)$ requires that the two diagonal blocks to be proportional to identity. Only one Dirac matrix out of the fifteen remains, and its coefficient is $(\epsilon_v-\epsilon_c)/2$. The band crossing appears at
\bea
\epsilon_v(k_x,k_y,\pi)=\epsilon_c(k_x,k_y,\pi)
\eea
which is one equation with two variables: the solution space is generically one-dimensional, \emph{i. e.}, a nodal line. Since both crossing bands are doubly degenerate, we call this crossing a double-nodal line.

The topological invariant for the double-nodal lines is very similar to the case of single nodal lines protected by mirror reflection. Choose two points on the two sides of the double-nodal line, $p_1$ and $p_2$, and count the number of the occupied bands at $p_{1,2}$ that have $M'=+i$, and denote them by $N_1$ and $N_2$. The $Z$-invariant is given by
\bea
Z=N_1-N_2.
\eea

We have one technical comment before closing this subsection. To simplify our discussion, in Eq.(\ref{eq:8}) we choose the simplest form of the twofold rotation, which passes through the inversion center (origin). In fact, the screw axis can also be shifted to $x=a/4$ or $y=b/4$. In that case, the reflection symmetry $M'=P*R$ is not a mirror reflection but a glide reflection symmetry, because it also involves half-lattice translation along the $x$- or $y$-direction. This will slightly complicate the proof that the two subbands have the same $M'$ eigenvalue, but the statement itself remains unchanged\cite{CFang_NLSM_PRB}.

\section{Materials realization}
Since TNLSM was proposed in 2011~\cite{Burkov_Topological_nodal_semimetals_2011PRB}, there have been
many material proposals to realize it experimentally. According to the classification in the former section, we to summarize, 
to the best of our knowledge, the existing proposals in Table.~\ref{nlsm_proposal}. The original proposal of 
Burkov {\it et al.}~\cite{Burkov_Topological_nodal_semimetals_2011PRB}
is based on a fine tuned superlattice of normal insulator and topological insulator with broken time reversal symmetry, which is actually 
a model rather than a realistic material. The first type-A NLSM was proposed in the half-metallic double WSM 
HgCr$_2$Se$_4$.~\cite{XuGang_HgCrSe_2011_PRL} When the magnetization is long [001] axis, there is a nodal 
line inside of the $k_z$=0 mirror plane in addition to the double Weyl nodes on the [001] axis\cite{Fang2012}. Such spinful NLSM 
has also been proposed in noncentrosymmetric TlTaSe$_2$~\cite{TlTaSe2} and PbTaSe$_2$~\cite{PbTaSe2}, where 
the bands are non-degenerate due to inversion symmetry breaking though time reversal symmetry is conserved. For spineless case, TaAs,~\cite{TaAs_Weng, HuangSM_Weyl} 
ZrTe~\cite{ZrTe} and CaAg$X$ ($X$=P, As)~\cite{CaAgX} have nodal line as protected by the mirror symmetry. In 
the first two materials, the nodal line decay into Weyl points when SOC is included while the last one becomes a 
fully gaped TI. In 2014, a kind of all carbon graphene network, so called Mackay-Terrones crystal (MTC)\cite{allcarbon_nodeLine2014}, 
was proposed to be NLSM of type B and the Bernal graphite~\cite{GP_Mikitik_2006PRB,GP_Mikitik_2008LTP} studied 
a decade ago was also revealed to be of this type. The other proposals for type B, including 
Ca$_3$P$_2$,~\cite{Ca3P2, Ca3P2-2} Cu$_3$(Pd,Zn)N,~\cite{Cu3NPd_Kane, Cu3NPd_Yu} 
LaN,~\cite{LnX} Be and other alkaline-earth metal,~\cite{Murakami_CaSrYb, ChenXQ_Be} 
CaTe,~\cite{CaTe} BaSn$_2$,~\cite{BaSn2} Black Phosphorus (BP) under pressure~\cite{BP_Zhao2015},
CaP$_3$~\cite{CaP3} and other carbon based materials like interpenetrated graphene network (IGN)~\cite{carbon_fanzhang} and 
body-centered orthorhombic C$_{16}$ (BCO-C16).~\cite{BCO-C16} Among them, CaP$_3$ 
has the lowest crystal symmetry and the nodal line appears at arbitrary points in the momentum space, 
while the others have their nodal line(s) constrained in the mirror plane(s). SOC can be neglected for 
compounds composed of light elements. Presently, only Be metal with very 
tiny SOC has been confirmed to host nodal line by the ARPES experiment.~\cite{ChenXQ_Be} 
For IGN in Ref.~\onlinecite{carbon_fanzhang}, the nodal line stretches over the whole BZ and connects to its image in the next BZ
instead of forming a closed ring. For type C, SrIrO$_3$~\cite{CFang_NLSM_PRB} and 
Ba$MX_3$ ($M$=V, Nb and Ta, $X$=S, Se)~\cite{BaMX3} have been proposed. The later one has been 
shown to host nodal surface when SOC is not taken into account. 
Different from the inner connecting of the three mutually perpendicular nodal rings in MTC and Cu$_3$(Pd, Zn)N,
the inter connecting of nodal rings can form nodal chain state, which has been proposed for IrF$_4$ family compounds.~\cite{nodal_chain} 

\begin{table*}

\caption{
 \label{nlsm_proposal}
The proposed materials to host nodal lines classified. The DSM, WSM and TI mean the nodal lines evolve into the corresponding topological state when spin-orbit coupling (SOC) is further included. N/A means unknown.}
\begin{tabular*}{\textwidth}{@{\extracolsep{\fill}}c|c|c}
\hline\hline
Class   & NO SOC  &  +SOC  \\
 	\hline
Type A & TaTa, ZrTe 				& Weyl semimetal \\
            & CaAg$X$ ($X$=P, As) 	& Topological insulators\\
            &						& HgCr$_2$Se$_4$, TlTaSe$_2$, PbTaSe$_2$\\
 	\hline
Type B 	& CaP$_3$ & Topological insulators \\
		& MTC, BaSn$_2$, BP, IGN, BCO-C$_{16}$	& Topological insulators\\
		& Be and other alkaline-earth metal, Ca$_3$P$_2$ & N/A\\
		&Cu$_3$(Pd, Zn)N, LaN, CaTe	& Dirac semimetals\\
 	\hline
Type C &  & SrIrO$_3$, Ba$M$$X_3$ ($M$=V, Nb, Ta, $X$=S, Se) \\
 	\hline\hline
\end{tabular*}
\end{table*}

\section{Symmetry breaking scenarios}

We have established several classes of nodal lines protected by different symmetry groups. When these symmetries are partially or fully broken, the nodal lines generically vanish with them. However, unlike a point node that can simply gap out, there are at least two fates for a nodal line: it may break into several discrete nodal points, or it may also be fully gapped. In this section, we show (i) how a nodal line protected by mirror reflection and SU(2) spin rotation breaks into several Weyl points or Dirac points if SU(2) is broken and (ii) how the double-nodal line proposed in SrIrO$_3$ breaks into two nonsymmorphic Dirac points when one glide reflection symmetry is broken.

\subsection{Nodal line broken into point nodes by SOC}

As we have discussed at the end of Sec.II, in a spinless (or SU(2)-symmetric) system, nodal line appears when there is a band inversion between two bands with opposite mirror eigenvalues. This has been seen in first principles calculations in many classes of materials. In reality, the spin-orbital coupling, while small, cannot be completely ignored in electronic systems. When 
SOC is considered, the band repulsion between opposite spins becomes nonzero at finite momenta, making the nodal line unstable. Depending on the remaining symmetries (other than SU(2)) of our system, the nodal line can either break into several pairs of Weyl points, one or several Dirac points, two separate nodal lines, or become fully gapped. In TaAs and several materials of the same family, it is found that a nodal line in the non-SOC band structure breaks into three pairs of Weyl points; in LaN, Cu$_3$(Pd,Zn)N and CaTe, three intersecting nodal lines break into two Dirac points. In noncentrosymmetric materials, a nodal line in the non-SOC band structure usually signifies Weyl points when SOC is turned on.

How a nodal line breaks into Weyl points can be understood in simple effective models, but the quantitative results (say how many pairs of Weyl points and their location) will differ from those from the first principle calculations. Consider an effective Hamiltonian for TNLSM protected by mirror reflection and SU(2)
\bea
H_0=(m-k^2)\sigma_zs_0+k_z\sigma_ys_0,
\eea
where $\sigma_{0,i}$ acts on the orbital index (for example, one orbital maybe $s$-orbital and the other $p_z$), and $s_{0,i}$ acts on the spin. Note that here only $s_0$ appears due to SU(2) symmetry. The symmetries are represented by $M=i\sigma_zs_z$ and $T=K(is_y)$. When SU(2) is broken, we can add spin-orbital terms
\bea
H=H_0-m'\sigma_ys_x+k_x\sigma_xs_x,
\eea
whose dispersion takes the form
\bea
E(\bk)=\sqrt{(m-k^2)^2+(k_z\pm{m'})^2+k_x^2}.
\eea
For $m>m'^2$, this dispersion has two positive Weyl points at
\bea
W_{1+}=(0,\sqrt{m-m'^2},m'),\\
\nonumber
W_{2+}=(0,-\sqrt{m-m'^2},-m')
\eea
and two negative Weyl points at
\bea
W_{1-}=(0,-\sqrt{m-m'^2},m'),\\
\nonumber
W_{2-}=(0,\sqrt{m-m'^2},-m').
\eea
At the point $m=m'^2$, $W_{i+}$ and $W_{i-}$ annihilate each other, and the system becomes fully gapped.

In the second example, we have mirror reflection symmetry about the $xz$- and $yz$-planes, and we assume there is fourfold rotation symmetry about the $z$-axis and inversion symmetry
\bea\label{eq:9}
H_0=(m-k^2)\sigma_zs_0+k_xk_y\sigma_xs_0,
\eea
and the symmetries are represented by $C_4=\sigma_z\exp(-is_z\pi/4)$, $M_{xz}=i\sigma_zs_y$, $M_{yz}=i\sigma_zs_x$ and $T=K(is_y)$. Solving for the energy dispersion of Eq.(\ref{eq:9}) we see that there are two nodal lines, which are the intersection between the $k_xk_z$-plane and the sphere of $k=\sqrt{m}$ and the $k_yk_z$-plane and the sphere. Then we add the SOC terms
\bea
H=H_0+k_y\sigma_xs_y-k_x\sigma_xs_x,
\eea
and the dispersion becomes
\bea
E(\bk)=\sqrt{(m-k^2)^2+(\sqrt{k_x^2+k_y^2}\pm{k_xk_y})^2}.
\eea
This dispersion only has two point nodes at $\bk=(0,0,\pm\sqrt{m})$, and at each point, all four bands meet at the same point, \emph{i. e.}, the two points are Dirac points.

\subsection{Double nodal line broken into nonsymmorphic Dirac points}

SrIrO$_3$ is the first proposed material that has a double-nodal line, protected in this particular case by a twofold screw rotation about the $b$-axis $S_y: (x,y,z)\rightarrow(-x+a/2,y+b/2,-z+c/2)$ as well as inversion and time-reversal symmetries. The space group of the bulk material is generated by inversion $P: (x,y,z)\rightarrow(-x,-y,-z)$, $S_y$ and $S_x: (x,y,z)\rightarrow(x+a/2,-y+b/2,-z)$. Now suppose we break $S_y$ preserving $P$ and $S_x$, the double-nodal line is no longer protected. However, one can prove that along $UR$, the subbands in a degenerate bands have the \emph{same} eigenvalues of $S_x$. To see this, notice that
\bea\label{eq:10}
S_x^2=-T_{100}=-e^{-ik_x},
\eea
where the minus sign comes from the full rotation of the $1/2$-spin. From Eq.(\ref{eq:10}), we find that the eigenvalues of $S_x$ to be $\pm{i}e^{-ik_x/2}$. Then consider the commutation relation between $P$ and $S_x$
\bea
S_x*P=T_{110}P*S_x=e^{-ik_x-ik_y}P*S_x.
\eea
Suppose $|\phi(\bk)\rangle$ is one subband of a degenerate band along $UR$ of $S_x$-eigenvalue $+ie^{-ik_x/2}$, then for the other subband
\bea\label{eq:11}
S_x(P*T|\phi(\bk)\rangle)&=&-e^{-ik_x}P*T*S_x|\phi(\bk)\rangle\\
\nonumber&=&-e^{-ik_x}P*T(+ie^{-ik_x/2})|\phi(\bk)\rangle\\
\nonumber
&=&ie^{-ik_x/2}P*T|\phi(\bk)\rangle.
\eea
Eq.(\ref{eq:11}) shows that, as promised, that the two subbands have the \emph{same} eigenvalue of $S_x$. Therefore, if two degenerate bands have opposite $S_x$-eigenvalues, they may cross each other at a Dirac point. The Dirac points protected by the twofold screw axis and inversion are distinct from normal Dirac semimetals in that they have \emph{topologically protected} surface states (double-helicoid surface states in this case).

\section{Physical consequences}

For most topological materials, the observation of topological surface states has been considered a definitive  confirmation of the nontrivial topology in the band structure. The underpinning of this logic is the bulk-edge correspondence principle, which asserts that a nontrivial bulk topology in $d$-dimensional bulk must correspond to a gapless mode $d-1$-dimensional edge, which cannot be realized in a real $d-1$-dimensional system without symmetry breaking. Here we emphasize the prerequisites of its application that (i) the symmetry group protecting the topology in the bulk must be unbroken on the edge and (ii) the interaction is weak or the edge can still be gapped into an anomalous topology order.

So far, the protection of nodal lines requires one or several of the following spatial symmetries: mirror reflection, space inversion and twofold screw rotation. A simple inspection of them shows that there is no surface where any of these symmetries is preserved. Therefore, the bulk-edge correspondence may not be applied here to indicate the existence of topological surface modes in TNLSMs. In numerical simulations, however, there are indeed states localized on the surface in the surface BZ, appearing inside the projections of the nodal lines. Unlike the surface states of, say, topological insulators, these surface states are very flat in dispersion, and are hence dubbed `drumhead' states. The drumhead states can be considered an higher-dimensional analogy of the flat band on the zigzag edge of graphene. As we have stated, the flat bands are \emph{not} topologically protected: a change of the model parameters on the surface will destroy the `flatness' of the surface modes, and can even push these surface states into the bulk continuum spectrum, the same way the flat band on the zigzag edge gains dispersion as soon as we turn on the intra-sublattice hopping on the edge. The lack of a topological signature on the surface poses a challenge to designing a `smoking gun' experimental observation of TNLSM. In Ref.[\onlinecite{TlTaSe2,PbTaSe2}], using ARPES, the group measure the dispersion of both the bulk and the surface states, where the results support the existence of a nodal line protected by mirror reflection symmetry in PbTaSe$_2$ and TlTaSe$_2$.

Due to the lack of surface signatures in TNLSMs, people turn to the bulk probes. In quantum oscillation, the special behavior of these materials have been predicted and measured. In Ref.[\onlinecite{Rhim2015}], the Landau levels of an effective $k\cdot{p}$-model near a double-nodal line is calculated as a function of the strength and the angle of the magnetic field. It is predicted that there are zero modes in the spectrum that lead to a peak in density of states at the Fermi level. In Ref.[\onlinecite{Mullen2015}], the Landau levels are calculated in a lattice model describing a 3D honeycomb lattice that has nodal rings in the BZ. A key distinction between this work and Ref.[] is that here the magnetic field is in the toroidal direction, where the field lines are parallel to the nodal ring. It is found that the Hall conductance, $\sigma_{z\rho}$ where $\rho$ means the radial direction in cylindrical coordinates, is quantized so that a 3D quantum Hall effect can be realized. In Ref.[\onlinecite{Hu2016}], the de Haas-van Alphen oscillation is measured in TNLSMs ZrSiSe and ZrSiTe. The authors use the angle-dependent oscillation frequencies to map out the Fermi surfaces in these two materials and by fitting the Lifshitz-Kosevich formula Berry phase is found to be $\sim0.31\pi$ for out-of-plane and $\sim\pi$ for in-plane field (The latter value matches the theoretical prediction.)

The ideal Fermi surface of a TNLSM is the nodal ring itself at half-filling, which may be achieved only if there is particle-hole symmetry or chiral symmetry that pins the energy of all the points on the nodal line at the Fermi energy. The dispersion near the Fermi surface is also particular in a TNLSM: while the band splitting along the nodal line is zero, the dispersion perpendicular to the nodal line is linear in momentum. These special properties near the Fermi energy lead to new many-body effects for TNLSMs. In Ref.[\onlinecite{Huh2016}], it is predicted that the Coulomb interaction is only partially screened and still long-ranged due to the vanishing density of states at the Fermi energy. In Ref.[\onlinecite{Han2016}], the authors using renormalization group method analyze the quantum phase transition between a topological nodal line superconductor and a fully gapped superconductor where the symmetry protecting the nodal line is broken. It is predicted that this transition belongs to a new universality class, where the dynamic exponent, order parameter exponent and susceptibility exponent are all $1$.

\section{Acknowledgments}

CF, HMW and XD are supported by the National Key Research and Development Program of China under grant No. 2016YFA0302400 and No. 2016YFA0300604. HMW, ZF and XD are supported by National Natural Science Foundation of China (Grant Nos. 11274359 and 11422428), the National 973 program of China (Grant No. 2013CB921700), the ``Strategic Priority Research Program (B)" of the Chinese Academy of Sciences (Grant No. XDB07020100).


\begin{thebibliography}{49}
\expandafter\ifx\csname natexlab\endcsname\relax\def\natexlab#1{#1}\fi
\expandafter\ifx\csname bibnamefont\endcsname\relax
  \def\bibnamefont#1{#1}\fi
\expandafter\ifx\csname bibfnamefont\endcsname\relax
  \def\bibfnamefont#1{#1}\fi
\expandafter\ifx\csname citenamefont\endcsname\relax
  \def\citenamefont#1{#1}\fi
\expandafter\ifx\csname url\endcsname\relax
  \def\url#1{\texttt{#1}}\fi
\expandafter\ifx\csname urlprefix\endcsname\relax\def\urlprefix{URL }\fi
\providecommand{\bibinfo}[2]{#2}
\providecommand{\eprint}[2][]{\url{#2}}

\bibitem[{\citenamefont{Murakami}(2007)}]{Murakami2007}
\bibinfo{author}{\bibfnamefont{S.}~\bibnamefont{Murakami}},
  \bibinfo{journal}{New Journal of Physics} \textbf{\bibinfo{volume}{9}},
  \bibinfo{pages}{356} (\bibinfo{year}{2007}).

\bibitem[{\citenamefont{Wan et~al.}({2011})\citenamefont{Wan, Turner,
  Vishwanath, and Savrasov}}]{Wan2011}
\bibinfo{author}{\bibfnamefont{X.}~\bibnamefont{Wan}},
  \bibinfo{author}{\bibfnamefont{A.~M.} \bibnamefont{Turner}},
  \bibinfo{author}{\bibfnamefont{A.}~\bibnamefont{Vishwanath}},
  \bibnamefont{and} \bibinfo{author}{\bibfnamefont{S.~Y.}
  \bibnamefont{Savrasov}}, \bibinfo{journal}{Phys. Rev. B}
  \textbf{\bibinfo{volume}{{83}}}, \bibinfo{pages}{205101}
  (\bibinfo{year}{{2011}}).

\bibitem[{\citenamefont{Wang et~al.}({2012})\citenamefont{Wang, Sun, Chen,
  Franchini, Xu, Weng, Dai, and Fang}}]{Wang2012}
\bibinfo{author}{\bibfnamefont{Z.}~\bibnamefont{Wang}},
  \bibinfo{author}{\bibfnamefont{Y.}~\bibnamefont{Sun}},
  \bibinfo{author}{\bibfnamefont{X.-Q.} \bibnamefont{Chen}},
  \bibinfo{author}{\bibfnamefont{C.}~\bibnamefont{Franchini}},
  \bibinfo{author}{\bibfnamefont{G.}~\bibnamefont{Xu}},
  \bibinfo{author}{\bibfnamefont{H.}~\bibnamefont{Weng}},
  \bibinfo{author}{\bibfnamefont{X.}~\bibnamefont{Dai}}, \bibnamefont{and}
  \bibinfo{author}{\bibfnamefont{Z.}~\bibnamefont{Fang}},
  \bibinfo{journal}{Phys. Rev. B} \textbf{\bibinfo{volume}{{85}}},
  \bibinfo{pages}{195320} (\bibinfo{year}{{2012}}), ISSN
  \bibinfo{issn}{{1098-0121}}.

\bibitem[{\citenamefont{Burkov et~al.}(2011)\citenamefont{Burkov, Hook, and
  Balents}}]{Burkov_Topological_nodal_semimetals_2011PRB}
\bibinfo{author}{\bibfnamefont{A.~A.} \bibnamefont{Burkov}},
  \bibinfo{author}{\bibfnamefont{M.~D.} \bibnamefont{Hook}}, \bibnamefont{and}
  \bibinfo{author}{\bibfnamefont{L.}~\bibnamefont{Balents}},
  \bibinfo{journal}{Phys. Rev. B} \textbf{\bibinfo{volume}{84}},
  \bibinfo{pages}{235126} (\bibinfo{year}{2011}).

\bibitem[{\citenamefont{Chen et~al.}(2015{\natexlab{a}})\citenamefont{Chen,
  Xie, Yang, Pan, Zhang, Cohen, and Zhang}}]{Chen2015}
\bibinfo{author}{\bibfnamefont{Y.}~\bibnamefont{Chen}},
  \bibinfo{author}{\bibfnamefont{Y.}~\bibnamefont{Xie}},
  \bibinfo{author}{\bibfnamefont{S.~A.} \bibnamefont{Yang}},
  \bibinfo{author}{\bibfnamefont{H.}~\bibnamefont{Pan}},
  \bibinfo{author}{\bibfnamefont{F.}~\bibnamefont{Zhang}},
  \bibinfo{author}{\bibfnamefont{M.~L.} \bibnamefont{Cohen}}, \bibnamefont{and}
  \bibinfo{author}{\bibfnamefont{S.}~\bibnamefont{Zhang}},
  \bibinfo{journal}{Nano. Lett.} \textbf{\bibinfo{volume}{15}},
  \bibinfo{pages}{6974} (\bibinfo{year}{2015}{\natexlab{a}}).

\bibitem[{\citenamefont{Xu et~al.}(2011)\citenamefont{Xu, Weng, Wang, Dai, and
  Fang}}]{XuGang_HgCrSe_2011_PRL}
\bibinfo{author}{\bibfnamefont{G.}~\bibnamefont{Xu}},
  \bibinfo{author}{\bibfnamefont{H.~M.} \bibnamefont{Weng}},
  \bibinfo{author}{\bibfnamefont{Z.~J.} \bibnamefont{Wang}},
  \bibinfo{author}{\bibfnamefont{X.}~\bibnamefont{Dai}}, \bibnamefont{and}
  \bibinfo{author}{\bibfnamefont{Z.}~\bibnamefont{Fang}},
  \bibinfo{journal}{Phys. Rev. Lett.} \textbf{\bibinfo{volume}{107}},
  \bibinfo{pages}{186806} (\bibinfo{year}{2011}).

\bibitem[{\citenamefont{{Bzdu{\v s}ek} et~al.}(2016)\citenamefont{{Bzdu{\v
  s}ek}, {Wu}, {R{\"u}egg}, {Sigrist}, and {Soluyanov}}}]{nodal_chain}
\bibinfo{author}{\bibfnamefont{T.}~\bibnamefont{{Bzdu{\v s}ek}}},
  \bibinfo{author}{\bibfnamefont{Q.}~\bibnamefont{{Wu}}},
  \bibinfo{author}{\bibfnamefont{A.}~\bibnamefont{{R{\"u}egg}}},
  \bibinfo{author}{\bibfnamefont{M.}~\bibnamefont{{Sigrist}}},
  \bibnamefont{and} \bibinfo{author}{\bibfnamefont{A.~A.}
  \bibnamefont{{Soluyanov}}}, \bibinfo{journal}{Nature}
  (\bibinfo{year}{2016}).

\bibitem[{\citenamefont{Lin et~al.}(2016)\citenamefont{Lin, Hu, Chen, Lee, and
  Zhang}}]{Lin2016}
\bibinfo{author}{\bibfnamefont{J.~Y.} \bibnamefont{Lin}},
  \bibinfo{author}{\bibfnamefont{N.~C.} \bibnamefont{Hu}},
  \bibinfo{author}{\bibfnamefont{Y.~J.} \bibnamefont{Chen}},
  \bibinfo{author}{\bibfnamefont{C.~H.} \bibnamefont{Lee}}, \bibnamefont{and}
  \bibinfo{author}{\bibfnamefont{X.}~\bibnamefont{Zhang}},
  \bibinfo{journal}{arXiv:1607.06524}  (\bibinfo{year}{2016}).

\bibitem[{\citenamefont{Chiu and Schnyder}(2014)}]{Chiu2014}
\bibinfo{author}{\bibfnamefont{C.-K.} \bibnamefont{Chiu}} \bibnamefont{and}
  \bibinfo{author}{\bibfnamefont{A.~P.} \bibnamefont{Schnyder}},
  \bibinfo{journal}{Phys. Rev. B} \textbf{\bibinfo{volume}{90}},
  \bibinfo{pages}{205136} (\bibinfo{year}{2014}),
  \urlprefix\url{http://link.aps.org/doi/10.1103/PhysRevB.90.205136}.

\bibitem[{\citenamefont{Yang et~al.}(2016)\citenamefont{Yang, Bojesen,
  Morimoto, and Furusaki}}]{Yang2016}
\bibinfo{author}{\bibfnamefont{B.-J.} \bibnamefont{Yang}},
  \bibinfo{author}{\bibfnamefont{T.~A.} \bibnamefont{Bojesen}},
  \bibinfo{author}{\bibfnamefont{T.}~\bibnamefont{Morimoto}}, \bibnamefont{and}
  \bibinfo{author}{\bibfnamefont{A.}~\bibnamefont{Furusaki}},
  \bibinfo{journal}{arXiv:1604.00483}  (\bibinfo{year}{2016}).

\bibitem[{\citenamefont{{Bian} et~al.}(2015{\natexlab{a}})\citenamefont{{Bian},
  {Chang}, {Zheng}, {Velury}, {Xu}, {Neupert}, {Chiu}, {Sanchez}, {Belopolski},
  {Alidoust} et~al.}}]{TlTaSe2}
\bibinfo{author}{\bibfnamefont{G.}~\bibnamefont{{Bian}}},
  \bibinfo{author}{\bibfnamefont{T.-R.} \bibnamefont{{Chang}}},
  \bibinfo{author}{\bibfnamefont{H.}~\bibnamefont{{Zheng}}},
  \bibinfo{author}{\bibfnamefont{S.}~\bibnamefont{{Velury}}},
  \bibinfo{author}{\bibfnamefont{S.-Y.} \bibnamefont{{Xu}}},
  \bibinfo{author}{\bibfnamefont{T.}~\bibnamefont{{Neupert}}},
  \bibinfo{author}{\bibfnamefont{C.-K.} \bibnamefont{{Chiu}}},
  \bibinfo{author}{\bibfnamefont{D.~S.} \bibnamefont{{Sanchez}}},
  \bibinfo{author}{\bibfnamefont{I.}~\bibnamefont{{Belopolski}}},
  \bibinfo{author}{\bibfnamefont{N.}~\bibnamefont{{Alidoust}}},
  \bibnamefont{et~al.}, \bibinfo{journal}{arXiv:1508.07521}
  (\bibinfo{year}{2015}{\natexlab{a}}),
  \urlprefix\url{http://adsabs.harvard.edu/abs/2015arXiv150807521B}.

\bibitem[{\citenamefont{{Bian} et~al.}(2015{\natexlab{b}})\citenamefont{{Bian},
  {Chang}, {Sankar}, {Xu}, {Zheng}, {Neupert}, {Chiu}, {Huang}, {Chang},
  {Belopolski} et~al.}}]{PbTaSe2}
\bibinfo{author}{\bibfnamefont{G.}~\bibnamefont{{Bian}}},
  \bibinfo{author}{\bibfnamefont{T.-R.} \bibnamefont{{Chang}}},
  \bibinfo{author}{\bibfnamefont{R.}~\bibnamefont{{Sankar}}},
  \bibinfo{author}{\bibfnamefont{S.-Y.} \bibnamefont{{Xu}}},
  \bibinfo{author}{\bibfnamefont{H.}~\bibnamefont{{Zheng}}},
  \bibinfo{author}{\bibfnamefont{T.}~\bibnamefont{{Neupert}}},
  \bibinfo{author}{\bibfnamefont{C.-K.} \bibnamefont{{Chiu}}},
  \bibinfo{author}{\bibfnamefont{S.-M.} \bibnamefont{{Huang}}},
  \bibinfo{author}{\bibfnamefont{G.}~\bibnamefont{{Chang}}},
  \bibinfo{author}{\bibfnamefont{I.}~\bibnamefont{{Belopolski}}},
  \bibnamefont{et~al.}, \bibinfo{journal}{arXiv:1505.03069}
  (\bibinfo{year}{2015}{\natexlab{b}}),
  \urlprefix\url{http://adsabs.harvard.edu/abs/2015arXiv150503069B}.

\bibitem[{\citenamefont{Yamakage et~al.}(2016)\citenamefont{Yamakage, Yamakawa,
  Tanaka, and Okamoto}}]{CaAgX}
\bibinfo{author}{\bibfnamefont{A.}~\bibnamefont{Yamakage}},
  \bibinfo{author}{\bibfnamefont{Y.}~\bibnamefont{Yamakawa}},
  \bibinfo{author}{\bibfnamefont{Y.}~\bibnamefont{Tanaka}}, \bibnamefont{and}
  \bibinfo{author}{\bibfnamefont{Y.}~\bibnamefont{Okamoto}},
  \bibinfo{journal}{Journal of the Physical Society of Japan}
  \textbf{\bibinfo{volume}{85}}, \bibinfo{pages}{013708}
  (\bibinfo{year}{2016}),
  \urlprefix\url{http://dx.doi.org/10.7566/JPSJ.85.013708}.

\bibitem[{\citenamefont{{Weng} et~al.}(2016)\citenamefont{{Weng}, {Fang},
  {Fang}, and {Dai}}}]{ZrTe}
\bibinfo{author}{\bibfnamefont{H.}~\bibnamefont{{Weng}}},
  \bibinfo{author}{\bibfnamefont{C.}~\bibnamefont{{Fang}}},
  \bibinfo{author}{\bibfnamefont{Z.}~\bibnamefont{{Fang}}}, \bibnamefont{and}
  \bibinfo{author}{\bibfnamefont{X.}~\bibnamefont{{Dai}}},
  \bibinfo{journal}{ArXiv e-prints}  (\bibinfo{year}{2016}),
  \eprint{1605.05186}.

\bibitem[{\citenamefont{Weng et~al.}(2015{\natexlab{a}})\citenamefont{Weng,
  Liang, Xu, Yu, Fang, Dai, and Kawazoe}}]{allcarbon_nodeLine2014}
\bibinfo{author}{\bibfnamefont{H.}~\bibnamefont{Weng}},
  \bibinfo{author}{\bibfnamefont{Y.}~\bibnamefont{Liang}},
  \bibinfo{author}{\bibfnamefont{Q.}~\bibnamefont{Xu}},
  \bibinfo{author}{\bibfnamefont{R.}~\bibnamefont{Yu}},
  \bibinfo{author}{\bibfnamefont{Z.}~\bibnamefont{Fang}},
  \bibinfo{author}{\bibfnamefont{X.}~\bibnamefont{Dai}}, \bibnamefont{and}
  \bibinfo{author}{\bibfnamefont{Y.}~\bibnamefont{Kawazoe}},
  \bibinfo{journal}{Phys. Rev. B} \textbf{\bibinfo{volume}{92}},
  \bibinfo{pages}{045108} (\bibinfo{year}{2015}{\natexlab{a}}),
  \urlprefix\url{http://link.aps.org/doi/10.1103/PhysRevB.92.045108}.

\bibitem[{\citenamefont{Xie et~al.}(2015)\citenamefont{Xie, Schoop, Seibel,
  Gibson, Xie, and Cava}}]{Ca3P2}
\bibinfo{author}{\bibfnamefont{L.~S.} \bibnamefont{Xie}},
  \bibinfo{author}{\bibfnamefont{L.~M.} \bibnamefont{Schoop}},
  \bibinfo{author}{\bibfnamefont{E.~M.} \bibnamefont{Seibel}},
  \bibinfo{author}{\bibfnamefont{Q.~D.} \bibnamefont{Gibson}},
  \bibinfo{author}{\bibfnamefont{W.}~\bibnamefont{Xie}}, \bibnamefont{and}
  \bibinfo{author}{\bibfnamefont{R.~J.} \bibnamefont{Cava}},
  \bibinfo{journal}{APL Mater.} \textbf{\bibinfo{volume}{3}},
  \bibinfo{eid}{083602} (\bibinfo{year}{2015}),
  \urlprefix\url{http://scitation.aip.org/content/aip/journal/aplmater/3/8/10.1063/1.4926545}.

\bibitem[{\citenamefont{{Chan} et~al.}(2015)\citenamefont{{Chan}, {Chiu},
  {Chou}, and {Schnyder}}}]{Ca3P2-2}
\bibinfo{author}{\bibfnamefont{Y.-H.} \bibnamefont{{Chan}}},
  \bibinfo{author}{\bibfnamefont{C.-K.} \bibnamefont{{Chiu}}},
  \bibinfo{author}{\bibfnamefont{M.~Y.} \bibnamefont{{Chou}}},
  \bibnamefont{and} \bibinfo{author}{\bibfnamefont{A.~P.}
  \bibnamefont{{Schnyder}}}, \bibinfo{journal}{ArXiv e-prints}
  (\bibinfo{year}{2015}), \eprint{1510.02759}.

\bibitem[{\citenamefont{Kim et~al.}(2015)\citenamefont{Kim, Wieder, Kane, and
  Rappe}}]{Cu3NPd_Kane}
\bibinfo{author}{\bibfnamefont{Y.}~\bibnamefont{Kim}},
  \bibinfo{author}{\bibfnamefont{B.~J.} \bibnamefont{Wieder}},
  \bibinfo{author}{\bibfnamefont{C.~L.} \bibnamefont{Kane}}, \bibnamefont{and}
  \bibinfo{author}{\bibfnamefont{A.~M.} \bibnamefont{Rappe}},
  \bibinfo{journal}{Phys. Rev. Lett.} \textbf{\bibinfo{volume}{115}},
  \bibinfo{pages}{036806} (\bibinfo{year}{2015}),
  \urlprefix\url{http://link.aps.org/doi/10.1103/PhysRevLett.115.036806}.

\bibitem[{\citenamefont{Yu et~al.}(2015)\citenamefont{Yu, Weng, Fang, Dai, and
  Hu}}]{Cu3NPd_Yu}
\bibinfo{author}{\bibfnamefont{R.}~\bibnamefont{Yu}},
  \bibinfo{author}{\bibfnamefont{H.}~\bibnamefont{Weng}},
  \bibinfo{author}{\bibfnamefont{Z.}~\bibnamefont{Fang}},
  \bibinfo{author}{\bibfnamefont{X.}~\bibnamefont{Dai}}, \bibnamefont{and}
  \bibinfo{author}{\bibfnamefont{X.}~\bibnamefont{Hu}}, \bibinfo{journal}{Phys.
  Rev. Lett.} \textbf{\bibinfo{volume}{115}}, \bibinfo{pages}{036807}
  (\bibinfo{year}{2015}),
  \urlprefix\url{http://link.aps.org/doi/10.1103/PhysRevLett.115.036807}.

\bibitem[{\citenamefont{{Zeng} et~al.}(2015)\citenamefont{{Zeng}, {Fang},
  {Chang}, {Chen}, {Hsieh}, {Bansil}, {Lin}, and {Fu}}}]{LnX}
\bibinfo{author}{\bibfnamefont{M.}~\bibnamefont{{Zeng}}},
  \bibinfo{author}{\bibfnamefont{C.}~\bibnamefont{{Fang}}},
  \bibinfo{author}{\bibfnamefont{G.}~\bibnamefont{{Chang}}},
  \bibinfo{author}{\bibfnamefont{Y.-A.} \bibnamefont{{Chen}}},
  \bibinfo{author}{\bibfnamefont{T.}~\bibnamefont{{Hsieh}}},
  \bibinfo{author}{\bibfnamefont{A.}~\bibnamefont{{Bansil}}},
  \bibinfo{author}{\bibfnamefont{H.}~\bibnamefont{{Lin}}}, \bibnamefont{and}
  \bibinfo{author}{\bibfnamefont{L.}~\bibnamefont{{Fu}}},
  \bibinfo{journal}{arXiv:1504.03492}  (\bibinfo{year}{2015}),
  \urlprefix\url{http://adsabs.harvard.edu/abs/2015arXiv150403492Z}.

\bibitem[{\citenamefont{{Hirayama} et~al.}(2016)\citenamefont{{Hirayama},
  {Okugawa}, {Miyake}, and {Murakami}}}]{Murakami_CaSrYb}
\bibinfo{author}{\bibfnamefont{M.}~\bibnamefont{{Hirayama}}},
  \bibinfo{author}{\bibfnamefont{R.}~\bibnamefont{{Okugawa}}},
  \bibinfo{author}{\bibfnamefont{T.}~\bibnamefont{{Miyake}}}, \bibnamefont{and}
  \bibinfo{author}{\bibfnamefont{S.}~\bibnamefont{{Murakami}}},
  \bibinfo{journal}{ArXiv e-prints}  (\bibinfo{year}{2016}),
  \eprint{1602.06501}.

\bibitem[{\citenamefont{{Li} et~al.}(2016)\citenamefont{{Li}, {Cheng}, {Ma},
  {Wang}, {Li}, {Zhang}, {Li}, and {Chen}}}]{ChenXQ_Be}
\bibinfo{author}{\bibfnamefont{R.}~\bibnamefont{{Li}}},
  \bibinfo{author}{\bibfnamefont{X.}~\bibnamefont{{Cheng}}},
  \bibinfo{author}{\bibfnamefont{H.}~\bibnamefont{{Ma}}},
  \bibinfo{author}{\bibfnamefont{S.}~\bibnamefont{{Wang}}},
  \bibinfo{author}{\bibfnamefont{D.}~\bibnamefont{{Li}}},
  \bibinfo{author}{\bibfnamefont{Z.}~\bibnamefont{{Zhang}}},
  \bibinfo{author}{\bibfnamefont{Y.}~\bibnamefont{{Li}}}, \bibnamefont{and}
  \bibinfo{author}{\bibfnamefont{X.-Q.} \bibnamefont{{Chen}}},
  \bibinfo{journal}{ArXiv e-prints}  (\bibinfo{year}{2016}),
  \eprint{1603.03974}.

\bibitem[{\citenamefont{{Du} et~al.}(2016)\citenamefont{{Du}, {Tang}, {Wang},
  {Sheng}, {Kan}, {Duan}, {Savrasov}, and {Wan}}}]{CaTe}
\bibinfo{author}{\bibfnamefont{Y.}~\bibnamefont{{Du}}},
  \bibinfo{author}{\bibfnamefont{F.}~\bibnamefont{{Tang}}},
  \bibinfo{author}{\bibfnamefont{D.}~\bibnamefont{{Wang}}},
  \bibinfo{author}{\bibfnamefont{L.}~\bibnamefont{{Sheng}}},
  \bibinfo{author}{\bibfnamefont{E.-j.} \bibnamefont{{Kan}}},
  \bibinfo{author}{\bibfnamefont{C.-G.} \bibnamefont{{Duan}}},
  \bibinfo{author}{\bibfnamefont{S.~Y.} \bibnamefont{{Savrasov}}},
  \bibnamefont{and} \bibinfo{author}{\bibfnamefont{X.}~\bibnamefont{{Wan}}},
  \bibinfo{journal}{ArXiv e-prints}  (\bibinfo{year}{2016}),
  \eprint{1605.07998}.

\bibitem[{\citenamefont{Huang et~al.}(2016)\citenamefont{Huang, Liu,
  Vanderbilt, and Duan}}]{BaSn2}
\bibinfo{author}{\bibfnamefont{H.}~\bibnamefont{Huang}},
  \bibinfo{author}{\bibfnamefont{J.}~\bibnamefont{Liu}},
  \bibinfo{author}{\bibfnamefont{D.}~\bibnamefont{Vanderbilt}},
  \bibnamefont{and} \bibinfo{author}{\bibfnamefont{W.}~\bibnamefont{Duan}},
  \bibinfo{journal}{Phys. Rev. B} \textbf{\bibinfo{volume}{93}},
  \bibinfo{pages}{201114} (\bibinfo{year}{2016}),
  \urlprefix\url{http://link.aps.org/doi/10.1103/PhysRevB.93.201114}.

\bibitem[{\citenamefont{{Zhao} et~al.}(2015)\citenamefont{{Zhao}, {Yu}, {Weng},
  and {Fang}}}]{BP_Zhao2015}
\bibinfo{author}{\bibfnamefont{J.}~\bibnamefont{{Zhao}}},
  \bibinfo{author}{\bibfnamefont{R.}~\bibnamefont{{Yu}}},
  \bibinfo{author}{\bibfnamefont{H.}~\bibnamefont{{Weng}}}, \bibnamefont{and}
  \bibinfo{author}{\bibfnamefont{Z.}~\bibnamefont{{Fang}}},
  \bibinfo{journal}{arXiv:1511.05704}  (\bibinfo{year}{2015}),
  \urlprefix\url{http://adsabs.harvard.edu/abs/2015arXiv151105704Z}.

\bibitem[{\citenamefont{{Xu} et~al.}(2016)\citenamefont{{Xu}, {Yu}, {Fang},
  {Dai}, and {Weng}}}]{CaP3}
\bibinfo{author}{\bibfnamefont{Q.}~\bibnamefont{{Xu}}},
  \bibinfo{author}{\bibfnamefont{R.}~\bibnamefont{{Yu}}},
  \bibinfo{author}{\bibfnamefont{Z.}~\bibnamefont{{Fang}}},
  \bibinfo{author}{\bibfnamefont{X.}~\bibnamefont{{Dai}}}, \bibnamefont{and}
  \bibinfo{author}{\bibfnamefont{H.}~\bibnamefont{{Weng}}},
  \bibinfo{journal}{ArXiv e-prints}  (\bibinfo{year}{2016}),
  \eprint{1608.03172}.

\bibitem[{\citenamefont{Chen et~al.}(2015{\natexlab{b}})\citenamefont{Chen,
  Xie, Yang, Pan, Zhang, Cohen, and Zhang}}]{carbon_fanzhang}
\bibinfo{author}{\bibfnamefont{Y.}~\bibnamefont{Chen}},
  \bibinfo{author}{\bibfnamefont{Y.}~\bibnamefont{Xie}},
  \bibinfo{author}{\bibfnamefont{S.~A.} \bibnamefont{Yang}},
  \bibinfo{author}{\bibfnamefont{H.}~\bibnamefont{Pan}},
  \bibinfo{author}{\bibfnamefont{F.}~\bibnamefont{Zhang}},
  \bibinfo{author}{\bibfnamefont{M.~L.} \bibnamefont{Cohen}}, \bibnamefont{and}
  \bibinfo{author}{\bibfnamefont{S.}~\bibnamefont{Zhang}},
  \bibinfo{journal}{Nano Letters} \textbf{\bibinfo{volume}{15}},
  \bibinfo{pages}{6974} (\bibinfo{year}{2015}{\natexlab{b}}),
  \urlprefix\url{http://dx.doi.org/10.1021/acs.nanolett.5b02978}.

\bibitem[{\citenamefont{Carter et~al.}(2012)\citenamefont{Carter, Shankar, Zeb,
  and Kee}}]{Carter2012}
\bibinfo{author}{\bibfnamefont{J.-M.} \bibnamefont{Carter}},
  \bibinfo{author}{\bibfnamefont{V.~V.} \bibnamefont{Shankar}},
  \bibinfo{author}{\bibfnamefont{M.~A.} \bibnamefont{Zeb}}, \bibnamefont{and}
  \bibinfo{author}{\bibfnamefont{H.-Y.} \bibnamefont{Kee}},
  \bibinfo{journal}{Phys. Rev. B} \textbf{\bibinfo{volume}{85}},
  \bibinfo{pages}{115105} (\bibinfo{year}{2012}).

\bibitem[{\citenamefont{Fang et~al.}(2015)\citenamefont{Fang, Chen, Kee, and
  Fu}}]{CFang_NLSM_PRB}
\bibinfo{author}{\bibfnamefont{C.}~\bibnamefont{Fang}},
  \bibinfo{author}{\bibfnamefont{Y.}~\bibnamefont{Chen}},
  \bibinfo{author}{\bibfnamefont{H.-Y.} \bibnamefont{Kee}}, \bibnamefont{and}
  \bibinfo{author}{\bibfnamefont{L.}~\bibnamefont{Fu}}, \bibinfo{journal}{Phys.
  Rev. B} \textbf{\bibinfo{volume}{92}}, \bibinfo{pages}{081201}
  (\bibinfo{year}{2015}),
  \urlprefix\url{http://link.aps.org/doi/10.1103/PhysRevB.92.081201}.

\bibitem[{\citenamefont{Liang et~al.}(2016)\citenamefont{Liang, Zhou, Yu, Wang,
  and Weng}}]{BaMX3}
\bibinfo{author}{\bibfnamefont{Q.-F.} \bibnamefont{Liang}},
  \bibinfo{author}{\bibfnamefont{J.}~\bibnamefont{Zhou}},
  \bibinfo{author}{\bibfnamefont{R.}~\bibnamefont{Yu}},
  \bibinfo{author}{\bibfnamefont{Z.}~\bibnamefont{Wang}}, \bibnamefont{and}
  \bibinfo{author}{\bibfnamefont{H.}~\bibnamefont{Weng}},
  \bibinfo{journal}{Phys. Rev. B} \textbf{\bibinfo{volume}{93}},
  \bibinfo{pages}{085427} (\bibinfo{year}{2016}),
  \urlprefix\url{http://link.aps.org/doi/10.1103/PhysRevB.93.085427}.

\bibitem[{\citenamefont{Weng et~al.}(2015{\natexlab{b}})\citenamefont{Weng,
  Fang, Fang, Bernevig, and Dai}}]{TaAs_Weng}
\bibinfo{author}{\bibfnamefont{H.}~\bibnamefont{Weng}},
  \bibinfo{author}{\bibfnamefont{C.}~\bibnamefont{Fang}},
  \bibinfo{author}{\bibfnamefont{Z.}~\bibnamefont{Fang}},
  \bibinfo{author}{\bibfnamefont{B.~A.} \bibnamefont{Bernevig}},
  \bibnamefont{and} \bibinfo{author}{\bibfnamefont{X.}~\bibnamefont{Dai}},
  \bibinfo{journal}{Phys. Rev. X} \textbf{\bibinfo{volume}{5}},
  \bibinfo{pages}{011029} (\bibinfo{year}{2015}{\natexlab{b}}),
  \urlprefix\url{http://link.aps.org/doi/10.1103/PhysRevX.5.011029}.

\bibitem[{\citenamefont{Huang et~al.}(2014)\citenamefont{Huang, Xu, Belopolski,
  Lee, Chang, Wang, Alidoust, Bian, Neupane, Zhang et~al.}}]{HuangSM_Weyl}
\bibinfo{author}{\bibfnamefont{S.~M.} \bibnamefont{Huang}},
  \bibinfo{author}{\bibfnamefont{S.~Y.} \bibnamefont{Xu}},
  \bibinfo{author}{\bibfnamefont{I.}~\bibnamefont{Belopolski}},
  \bibinfo{author}{\bibfnamefont{C.~C.} \bibnamefont{Lee}},
  \bibinfo{author}{\bibfnamefont{G.}~\bibnamefont{Chang}},
  \bibinfo{author}{\bibfnamefont{B.~K.} \bibnamefont{Wang}},
  \bibinfo{author}{\bibfnamefont{N.}~\bibnamefont{Alidoust}},
  \bibinfo{author}{\bibfnamefont{G.}~\bibnamefont{Bian}},
  \bibinfo{author}{\bibfnamefont{M.}~\bibnamefont{Neupane}},
  \bibinfo{author}{\bibfnamefont{C.}~\bibnamefont{Zhang}},
  \bibnamefont{et~al.}, \bibinfo{journal}{Nature Communications}
  \textbf{\bibinfo{volume}{6}}, \bibinfo{pages}{7373} (\bibinfo{year}{2014}).

\bibitem[{\citenamefont{Fang et~al.}(2016)\citenamefont{Fang, Lu, Liu, and
  Fu}}]{Fang2016}
\bibinfo{author}{\bibfnamefont{C.}~\bibnamefont{Fang}},
  \bibinfo{author}{\bibfnamefont{L.}~\bibnamefont{Lu}},
  \bibinfo{author}{\bibfnamefont{J.}~\bibnamefont{Liu}}, \bibnamefont{and}
  \bibinfo{author}{\bibfnamefont{L.}~\bibnamefont{Fu}},
  \bibinfo{journal}{Nature Physics}  (\bibinfo{year}{2016}).

\bibitem[{\citenamefont{Hu et~al.}(2016)\citenamefont{Hu, Tang, Liu, Liu, Zhu,
  Graf, Myhro, Tran, Lau, Wei et~al.}}]{Hu2016}
\bibinfo{author}{\bibfnamefont{J.}~\bibnamefont{Hu}},
  \bibinfo{author}{\bibfnamefont{Z.}~\bibnamefont{Tang}},
  \bibinfo{author}{\bibfnamefont{J.}~\bibnamefont{Liu}},
  \bibinfo{author}{\bibfnamefont{X.}~\bibnamefont{Liu}},
  \bibinfo{author}{\bibfnamefont{Y.}~\bibnamefont{Zhu}},
  \bibinfo{author}{\bibfnamefont{D.}~\bibnamefont{Graf}},
  \bibinfo{author}{\bibfnamefont{K.}~\bibnamefont{Myhro}},
  \bibinfo{author}{\bibfnamefont{S.}~\bibnamefont{Tran}},
  \bibinfo{author}{\bibfnamefont{C.~N.} \bibnamefont{Lau}},
  \bibinfo{author}{\bibfnamefont{J.}~\bibnamefont{Wei}}, \bibnamefont{et~al.},
  \bibinfo{journal}{Phys. Rev. Lett.} \textbf{\bibinfo{volume}{117}},
  \bibinfo{pages}{016602} (\bibinfo{year}{2016}),
  \urlprefix\url{http://link.aps.org/doi/10.1103/PhysRevLett.117.016602}.

\bibitem[{\citenamefont{Huh et~al.}(2016)\citenamefont{Huh, Moon, and
  Kim}}]{Huh2016}
\bibinfo{author}{\bibfnamefont{Y.}~\bibnamefont{Huh}},
  \bibinfo{author}{\bibfnamefont{E.-G.} \bibnamefont{Moon}}, \bibnamefont{and}
  \bibinfo{author}{\bibfnamefont{Y.~B.} \bibnamefont{Kim}},
  \bibinfo{journal}{Phys. Rev. B} \textbf{\bibinfo{volume}{93}},
  \bibinfo{pages}{035138} (\bibinfo{year}{2016}),
  \urlprefix\url{http://link.aps.org/doi/10.1103/PhysRevB.93.035138}.

\bibitem[{\citenamefont{Han et~al.}(2016)\citenamefont{Han, Cho, and
  Moon}}]{Han2016}
\bibinfo{author}{\bibfnamefont{S.}~\bibnamefont{Han}},
  \bibinfo{author}{\bibfnamefont{G.~Y.} \bibnamefont{Cho}}, \bibnamefont{and}
  \bibinfo{author}{\bibfnamefont{E.-G.} \bibnamefont{Moon}},
  \bibinfo{journal}{arXiv:1601.00975}  (\bibinfo{year}{2016}).

\bibitem[{\citenamefont{Haldane}(1988)}]{Haldane1988}
\bibinfo{author}{\bibfnamefont{F.~D.~M.} \bibnamefont{Haldane}},
  \bibinfo{journal}{Phys. Rev. Lett.} \textbf{\bibinfo{volume}{61}},
  \bibinfo{pages}{2015} (\bibinfo{year}{1988}),
  \urlprefix\url{http://link.aps.org/doi/10.1103/PhysRevLett.61.2015}.

\bibitem[{\citenamefont{Kane and Mele}(2005{\natexlab{a}})}]{Kane2005}
\bibinfo{author}{\bibfnamefont{C.~L.} \bibnamefont{Kane}} \bibnamefont{and}
  \bibinfo{author}{\bibfnamefont{E.~J.} \bibnamefont{Mele}},
  \bibinfo{journal}{Phys. Rev. Lett.} \textbf{\bibinfo{volume}{95}},
  \bibinfo{pages}{146802} (\bibinfo{year}{2005}{\natexlab{a}}),
  \urlprefix\url{http://link.aps.org/doi/10.1103/PhysRevLett.95.146802}.

\bibitem[{\citenamefont{Fu et~al.}(2007)\citenamefont{Fu, Kane, and
  Mele}}]{Fu2007a}
\bibinfo{author}{\bibfnamefont{L.}~\bibnamefont{Fu}},
  \bibinfo{author}{\bibfnamefont{C.}~\bibnamefont{Kane}}, \bibnamefont{and}
  \bibinfo{author}{\bibfnamefont{E.}~\bibnamefont{Mele}},
  \bibinfo{journal}{Phys. Rev. Lett.} \textbf{\bibinfo{volume}{98}},
  \bibinfo{pages}{106803} (\bibinfo{year}{2007}).

\bibitem[{\citenamefont{Moore and Balents}(2007)}]{Moore2007}
\bibinfo{author}{\bibfnamefont{J.~E.} \bibnamefont{Moore}} \bibnamefont{and}
  \bibinfo{author}{\bibfnamefont{L.}~\bibnamefont{Balents}},
  \bibinfo{journal}{Phys. Rev. B} \textbf{\bibinfo{volume}{75}},
  \bibinfo{pages}{121306} (\bibinfo{year}{2007}),
  \urlprefix\url{http://link.aps.org/doi/10.1103/PhysRevB.75.121306}.

\bibitem[{\citenamefont{Kane and Mele}(2005{\natexlab{b}})}]{Kane2005a}
\bibinfo{author}{\bibfnamefont{C.~L.} \bibnamefont{Kane}} \bibnamefont{and}
  \bibinfo{author}{\bibfnamefont{E.~J.} \bibnamefont{Mele}},
  \bibinfo{journal}{Phys. Rev. Lett.} \textbf{\bibinfo{volume}{95}},
  \bibinfo{pages}{226801} (\bibinfo{year}{2005}{\natexlab{b}}),
  \urlprefix\url{http://link.aps.org/doi/10.1103/PhysRevLett.95.226801}.

\bibitem[{\citenamefont{Bernevig and Zhang}(2006)}]{Bernevig2006}
\bibinfo{author}{\bibfnamefont{B.~A.} \bibnamefont{Bernevig}} \bibnamefont{and}
  \bibinfo{author}{\bibfnamefont{S.-C.} \bibnamefont{Zhang}},
  \bibinfo{journal}{Phys. Rev. Lett.} \textbf{\bibinfo{volume}{96}},
  \bibinfo{pages}{106802} (\bibinfo{year}{2006}),
  \urlprefix\url{http://link.aps.org/doi/10.1103/PhysRevLett.96.106802}.

\bibitem[{\citenamefont{Fang et~al.}(2012)\citenamefont{Fang, Gilbert, Dai, and
  Bernevig}}]{Fang2012}
\bibinfo{author}{\bibfnamefont{C.}~\bibnamefont{Fang}},
  \bibinfo{author}{\bibfnamefont{M.~J.} \bibnamefont{Gilbert}},
  \bibinfo{author}{\bibfnamefont{X.}~\bibnamefont{Dai}}, \bibnamefont{and}
  \bibinfo{author}{\bibfnamefont{B.~A.} \bibnamefont{Bernevig}},
  \bibinfo{journal}{Phys. Rev. Lett.} \textbf{\bibinfo{volume}{108}},
  \bibinfo{pages}{266802} (\bibinfo{year}{2012}).

\bibitem[{\citenamefont{Wang et~al.}(2013)\citenamefont{Wang, Weng, Wu, Dai,
  and Fang}}]{Wang2013}
\bibinfo{author}{\bibfnamefont{Z.}~\bibnamefont{Wang}},
  \bibinfo{author}{\bibfnamefont{H.}~\bibnamefont{Weng}},
  \bibinfo{author}{\bibfnamefont{Q.}~\bibnamefont{Wu}},
  \bibinfo{author}{\bibfnamefont{X.}~\bibnamefont{Dai}}, \bibnamefont{and}
  \bibinfo{author}{\bibfnamefont{Z.}~\bibnamefont{Fang}},
  \bibinfo{journal}{Phys. Rev. B} \textbf{\bibinfo{volume}{88}},
  \bibinfo{pages}{125427} (\bibinfo{year}{2013}).

\bibitem[{\citenamefont{Mikitik and Sharlai}(2006)}]{GP_Mikitik_2006PRB}
\bibinfo{author}{\bibfnamefont{G.~P.} \bibnamefont{Mikitik}} \bibnamefont{and}
  \bibinfo{author}{\bibfnamefont{Y.~V.} \bibnamefont{Sharlai}},
  \bibinfo{journal}{Physical Review B} \textbf{\bibinfo{volume}{73}},
  \bibinfo{pages}{235112} (\bibinfo{year}{2006}).

\bibitem[{\citenamefont{Mikitik and Sharlai}(2008)}]{GP_Mikitik_2008LTP}
\bibinfo{author}{\bibfnamefont{G.~P.} \bibnamefont{Mikitik}} \bibnamefont{and}
  \bibinfo{author}{\bibfnamefont{Y.~V.} \bibnamefont{Sharlai}},
  \bibinfo{journal}{Low Temperature Physics} \textbf{\bibinfo{volume}{34}},
  \bibinfo{pages}{794} (\bibinfo{year}{2008}).

\bibitem[{\citenamefont{Wang et~al.}(2016)\citenamefont{Wang, Weng, Nie, Fang,
  Kawazoe, and Chen}}]{BCO-C16}
\bibinfo{author}{\bibfnamefont{J.-T.} \bibnamefont{Wang}},
  \bibinfo{author}{\bibfnamefont{H.}~\bibnamefont{Weng}},
  \bibinfo{author}{\bibfnamefont{S.}~\bibnamefont{Nie}},
  \bibinfo{author}{\bibfnamefont{Z.}~\bibnamefont{Fang}},
  \bibinfo{author}{\bibfnamefont{Y.}~\bibnamefont{Kawazoe}}, \bibnamefont{and}
  \bibinfo{author}{\bibfnamefont{C.}~\bibnamefont{Chen}},
  \bibinfo{journal}{Phys. Rev. Lett.} \textbf{\bibinfo{volume}{116}},
  \bibinfo{pages}{195501} (\bibinfo{year}{2016}),
  \urlprefix\url{http://link.aps.org/doi/10.1103/PhysRevLett.116.195501}.

\bibitem[{\citenamefont{Rhim and Kim}(2015)}]{Rhim2015}
\bibinfo{author}{\bibfnamefont{J.-W.} \bibnamefont{Rhim}} \bibnamefont{and}
  \bibinfo{author}{\bibfnamefont{Y.~B.} \bibnamefont{Kim}},
  \bibinfo{journal}{Phys. Rev. B} \textbf{\bibinfo{volume}{92}},
  \bibinfo{pages}{045126} (\bibinfo{year}{2015}),
  \urlprefix\url{http://link.aps.org/doi/10.1103/PhysRevB.92.045126}.

\bibitem[{\citenamefont{Mullen et~al.}(2015)\citenamefont{Mullen, Uchoa, and
  Glatzhofer}}]{Mullen2015}
\bibinfo{author}{\bibfnamefont{K.}~\bibnamefont{Mullen}},
  \bibinfo{author}{\bibfnamefont{B.}~\bibnamefont{Uchoa}}, \bibnamefont{and}
  \bibinfo{author}{\bibfnamefont{D.~T.} \bibnamefont{Glatzhofer}},
  \bibinfo{journal}{Phys. Rev. Lett.} \textbf{\bibinfo{volume}{115}},
  \bibinfo{pages}{026403} (\bibinfo{year}{2015}),
  \urlprefix\url{http://link.aps.org/doi/10.1103/PhysRevLett.115.026403}.

\end{thebibliography}
\end{document}